\documentclass[%
 reprint,
 amsmath,amssymb,
 aps,
floatfix,
]{revtex4-2}

\usepackage{graphicx}
\usepackage{dcolumn}
\usepackage{bm}
\usepackage{color}
\usepackage{float}

\begin{document}

\preprint{APS/123-QED}

\title{Analysis of temporal structure of laser chaos by Allan variance}

\author{Naoki Asuke$^1$}
\email{jintengzhizi196@gmail.com}
\author{Nicolas Chauvet$^1$}
\author{Andr\'{e} R\"{o}hm$^1$}
\author{Kazutaka Kanno$^2$}
\author{Atsushi Uchida$^2$}
\author{Tomoaki Niiyama$^3$}
\author{Satoshi Sunada$^3$}
\author{Ryoichi Horisaki$^1$}
\author{Makoto Naruse$^1$}
\affiliation{$^1$ Department of Information Physics and Computing, Graduate School of Information Science and Technology, The University of Tokyo, 7-3-1 Hongo, Bunkyo-ku, Tokyo 113-8656, Japan}

\affiliation{$^2$ Department of Information and Computer Sciences, Saitama University, 255 Shimo-Okubo, Sakura-ku, Saitama city, Saitama 338-8570, Japan}

\affiliation{$^3$ Faculty of Mechanical Engineering, Institute of Science and Engineering, Kanazawa University, Kakuma, Kanazawa city, Ishikawa 920-1192, Japan}

\date{\today}

\begin{abstract}
Allan variance has been widely utilized for evaluating the stability of the time series generated by atomic clocks and lasers, in time regimes ranging from short to extremely long. This multi-scale examination capability of the Allan variance may also be beneficial in evaluating the chaotic oscillating dynamics of semiconductor lasers--- not just for conventional phase stability analysis. In the present study, we demonstrated Allan-variance analysis of the complex time series generated by a semiconductor laser with delayed feedback, including low-frequency fluctuations (LFFs), which exhibit both fast and slow dynamics. While the detection of LFFs is difficult with the conventional power spectrum analysis method in the low-frequency regime, the Allan-variance approach clearly captured the appearance of multiple time-scale dynamics, such as LFFs. This study demonstrates that Allan variance can help in understanding and characterizing diverse laser dynamics, including LFFs, spanning a wide range of timescales.
\end{abstract}

\maketitle


\section{\label{sec:intro}Introduction}
Allan variance has been widely utilized for evaluating the stability of oscillators, such as atomic clocks~\cite{AV}, quartz crystals~\cite{crystal}, and lasers~\cite{laserAV}, at a variety of timescales ranging from short to long. The phase stability of clocks and lasers is particularly important from the viewpoint of reliability~\cite{AV,crystal,laserAV}. On the other hand, intensive studies have been conducted to utilize ultrafast irregular time series generated by lasers, and to achieve improved levels of performance in application systems. Examples include, but are not limited to, reservoir computing~\cite{Larger:12}, random number generation~\cite{Uchida2008,RNG1,Zhang:12}, secure communications~\cite{Argyris2005,Rogister:01,ChaosComm2}, and decision making~\cite{decision}. In these systems, irregularity, instability, or complexity of the time series generated by lasers is utilized rather than their stable operation.

We consider that the multi-scale examination capability of the Allan variance may be useful for evaluating the chaotic oscillating dynamics of semiconductor lasers. In particular, a semiconductor laser that is subjected to delayed feedback exhibits versatile oscillatory dynamics~\cite{DelayChaos1,CHO1986131}. Low-frequency fluctuations (LFFs) exhibit both fast and slow dynamics. In LFFs, in addition to chaotically oscillating fast dynamics, a sudden dropout in the output light intensity and a gradual recovery is observed~\cite{Uchida12}. Furthermore, such a dropout does not occur in a periodic manner; that is, LFFs contain multi-scale attributes in the time domain. In this study, we demonstrate Allan-variance analysis of a complex time series generated by a semiconductor laser, which is described by the Lang--Kobayashi equations, with delayed optical feedback, including LFFs. The detection of LFFs via the conventional Fourier transform approach has been difficult, hampering the definition and classification of LFFs~\cite{Yamasaki:21}. In the present study, the Allan-variance approach clearly captured LFFs in a stable manner. 

The remainder of this paper is organized as follows. We review Allan variance and its conventional usage in Section~\ref{sec:AV}. The notion of LFFs in lasers and relevant literature are reviewed in Section~\ref{sec:LFF}. Section~\ref{sec:simulation} examines the Allan variance of the chaotic oscillating dynamics of lasers, including LFFs, on the basis of time series described by the Lang--Kobayashi equations~\cite{LK1980}, which are model equations for a semiconductor laser subjected to delayed optical feedback. The power spectral density and Allan variance of the obtained time series are examined and compared with regard to the discriminability of the dynamics. Section~\ref{sec:discussion} discusses the advantages of the Allan variance and its physical interpretation. Section~\ref{sec:conclusion} concludes the paper.
\section{Theory}
\subsection{\label{sec:AV} Allan variance}

\begin{figure*}[htb]
    \centering
    \includegraphics[width=1.95\columnwidth]{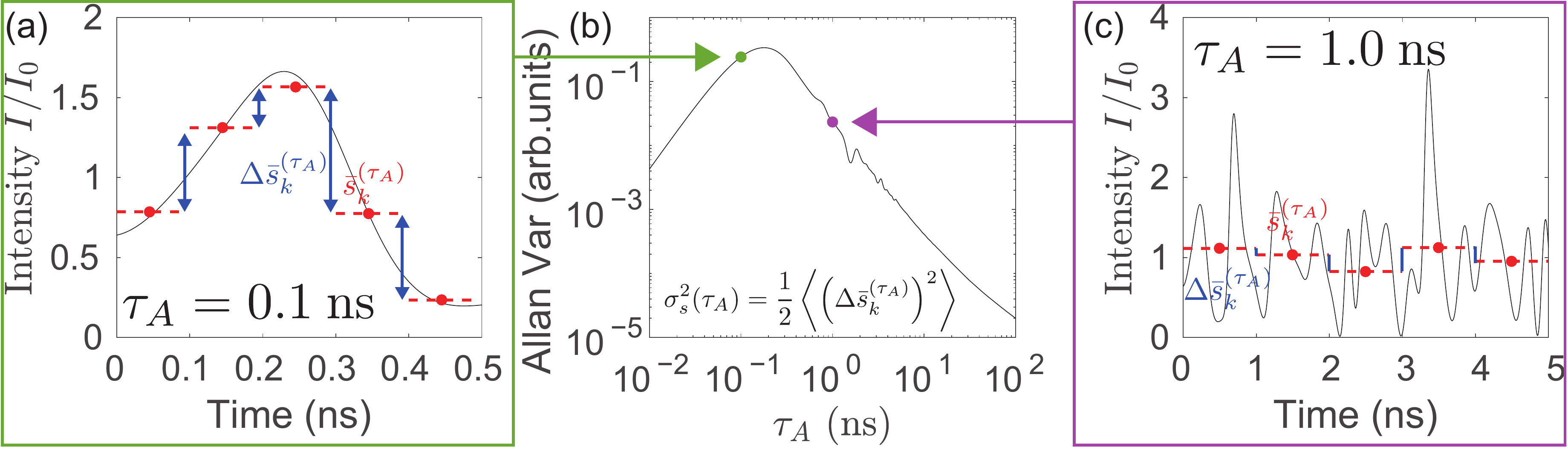}
\caption{
Allan variance of a chaotic laser time series. (a, c) Example of a laser chaos time series, normalized with respect to the intensity in the single mode. The Allan variance considers the variability of a time series for a given timescale ($\tau_A$). The red dashed lines indicate the mean values of the original time series, with $\tau_A$ being 0.1 ns in (a) and 1.0 ns in (c). The difference between consecutive timeslots, i.e., $\Delta\bar{s}_k^{(\tau_A)}$, is indicated by blue arrows. (b) Allan variance is defined as the variance of $\Delta\bar{s}_k^{(\tau_A)}$ with $\tau_A$ ranging from $10^{-2}$ to $10^2$ ns. Both axes are logarithmic. The green and purple dots represent the Allan variance when $\tau_A$ is 0.1 and 1.0 ns, respectively.}
    \label{fig:AV}
\end{figure*}

The Allan variance of a time series $s(t)$ is defined as
\begin{align}
    \sigma_s^2(\tau_A) &= \dfrac{1}{2}\left<\left(\Delta\bar{s}_k^{(\tau_A)}\right)^2\right>, \label{eq:AV_1} 
\end{align}
where 
\begin{align}
\Delta\bar{s}_k^{(\tau_A)} &= \bar{s}_k^{(\tau_A)} - \bar{s}_{k-1}^{(\tau_A)}, \label{eq:AV_2} \\
\bar{s}_k^{(\tau_A)} &= \dfrac{1}{\tau_A} \int_{t_k}^{t_{k+1} = t_k+\tau_A} s(t) dt, \label{eq:AV_3} \\
t_k &= t_0 + k\tau_A. \label{eq:AV_4}
\end{align}
Thus, the Allan variance considers the variance of the time series under study with respect to the timescale given by $\tau_A$. In Fig.~\ref{fig:AV}(a), the average of a time series over a period $\tau_A = 0.1$ ns is indicated by red dashed lines ($\bar{s}_k^{(\tau_A)}$), and the difference between successive slots is indicated by blue arrows ($\Delta\bar{s}_k^{(\tau_A)}$). The variance is indicated by the green point in Fig.~\ref{fig:AV}(b). 

The same time series is evaluated on a longer timescale, i.e., $\tau_A = 1.0$ ns, in Fig.~\ref{fig:AV}(c). The variance is indicated by the purple point in Fig.~\ref{fig:AV}(b). The Allan variance is indicated by the black curve in Fig.~\ref{fig:AV}(b), which captures the variability in various timescales ranging from a short timescale, which can correspond to the time resolution, to a long timescale that is almost the total recording time. Here, the time series and the calculated Allan variance in Fig.~\ref{fig:AV} are from the laser chaos time series, which is discussed in Sections~\ref{sec:simulation} and~\ref{sec:discussion}. 

It is known that the power spectral density of the frequency fluctuations of atomic clocks and lasers can be approximately expressed by the following equation~\cite{FSTABILITY}: 
\begin{align}
    S(f) &\approx \sum_{\alpha = -2}^{2} h_\alpha f^\alpha \label{eq:PhaseNoise_PSD}
\end{align}
where $h_\alpha$ denotes the coefficients for the approximation. Calculating these coefficients is equivalent to evaluating the phase stability.

The Allan variance can be expressed as follows using the power spectral density~\cite{FSTABILITY}: 
\begin{align}
    \sigma^2_s(\tau_A) &= 2\int_0^{\infty} S(f)\dfrac{\sin^4(\pi f \tau_A)}{(\pi f \tau_A)^2} df. \label{eq:AV_PSD}
\end{align}
Using Eq.~(\ref{eq:AV_PSD}), $h_\alpha$ can be obtained from the Allan variance of the phase time series data.
For example, for $\alpha = 0$, the Allan variance is 
\begin{align}
    2\int_0^\infty h_0\dfrac{\sin^4(\pi f \tau_A)}{(\pi f \tau_A)^2} df = \dfrac{h_0}{2\tau_A}. \label{eq:PhaseStabilityExample}
\end{align}
We can estimate $h_0$ from the Allan variance plot.
Thus, the Allan variance can be used to evaluate the variability characteristics of time series data.

In this study, we show that the Allan variance can be used to evaluate a chaotic time series from a new aspect, i.e., by applying it to a laser intensity time series instead of a phase time series.

\subsection{\label{sec:LFF} Low-frequency fluctuations (LFFs)}
\begin{figure}[htb]
    \centering
    \includegraphics[width=\columnwidth]{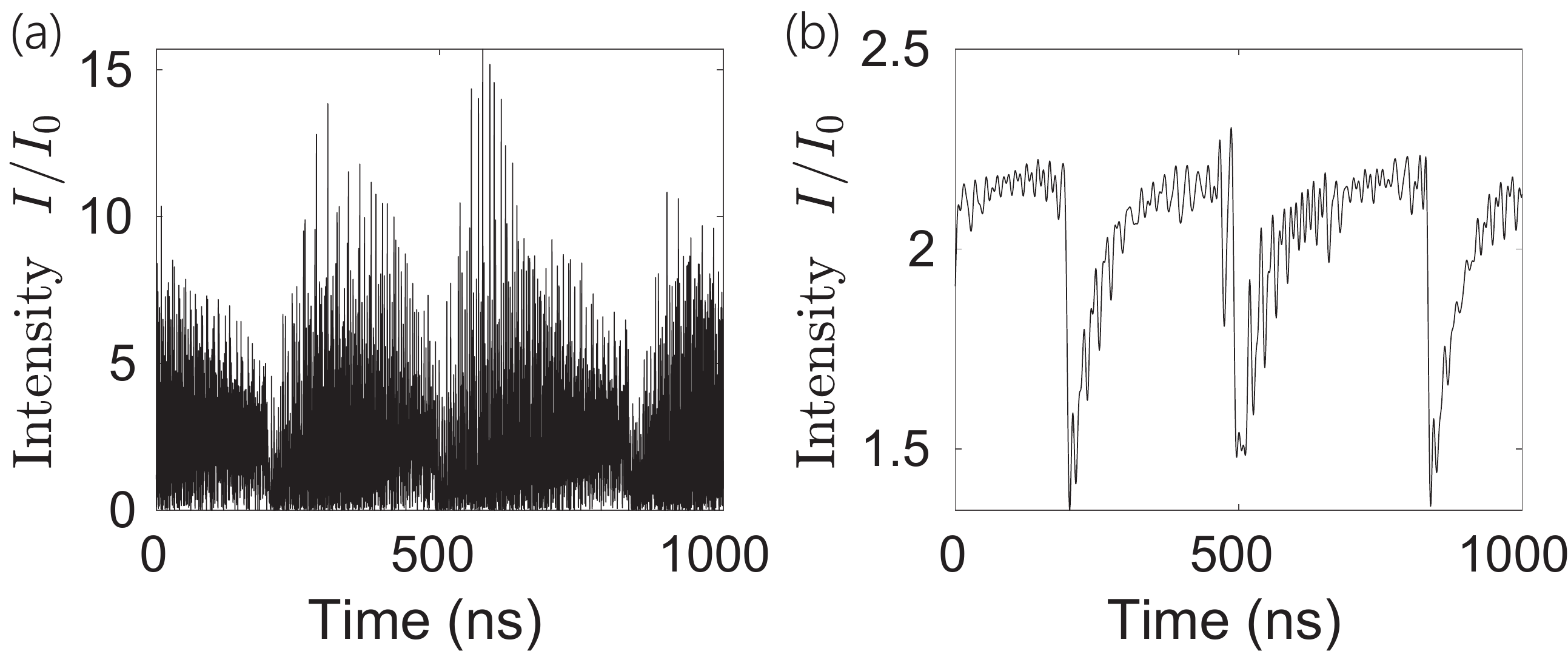}
    \caption{Example of an LFF time series. (a) Original time series. (b) Time series obtained by applying an ideal low-pass filter with a cutoff frequency of 100 MHz to the time series of (a). Sudden dropouts with gradual recovery are observed.}
    \label{fig:LFF}
\end{figure}

Semiconductor lasers are known to exhibit complex dynamics with the introduction of optical feedback~\cite{Ohtsubo17}, current modulation~\cite{Kao93}, etc. In recent years, these dynamics have been actively studied as the basis for new types of computing, such as reservoir computing~\cite{ORC}. LFFs are remarkable phenomena where GHz-order chaotic oscillations and MHz-order irregular dropouts and recoveries coexist. An example of an LFF time series is shown in Fig.~\ref{fig:LFF}. LFFs involve both fast and slow dynamics, which is a characteristic multi-scale complex phenomenon induced by delayed optical feedback. Furthermore, LFFs have been studied for application to random number generation~\cite{Bosco17} and decision making~\cite{Mihana:20}. However, a systematic and unified method for detecting and evaluating LFFs has not yet been established.

Yamasaki \textit{et al.}~\cite{Yamasaki:21} used the power spectrum density (PSD) to detect LFFs. However, it is technologically difficult to evaluate power spectral density accurately enough to stably distinguish LFFs from coherence collapse (CC), which only exhibits fast irregular fluctuations~\cite{CC}. Therefore, a new method that can automatically identify the dynamics is desirable for advancing basic research, evaluating optical devices, and establish system principles. 

In this study, we show that the Allan variance is useful for automatic LFF detection. This can be intuitively understood as follows. First, the Allan variance can be directly estimated from time-series data. Therefore, there is no need to evaluate the power spectral density. Second, the Allan variance can be used to capture and compare the variability characteristics in various timescales so that the fast chaotic oscillations and the slow dropouts exhibited by LFFs can be captured simultaneously, as shown in the next section.

\section{\label{sec:simulation} Allan variance analysis of laser chaos}
We generate a time series using the Lang--Kobayashi equations, which are model equations of semiconductor lasers with optical feedback and are given by as follows:
\begin{eqnarray}
    \dfrac{dE(t)}{dt} = \dfrac{1+i\alpha}{2}\left[\dfrac{G_N(N(t)-N_0)}{1+\varepsilon\|E(t)\|^2} - \dfrac{1}{T_P}\right]E(t) \nonumber \\ 
    + \kappa E(t-\tau_D)e^{-i\omega\tau_D}, \label{eq:LK-E} \\
    \dfrac{dN(t)}{dt} =
    J - \dfrac{N(t)}{T_S} - \dfrac{G_N(N(t)-N_0)\|E(t)\|^2}{1 + \varepsilon \|E(t)\|^2}, \label{eq:LK-N}
\end{eqnarray}
where $E(t)$ and $N(t)$ represent the complex electric field and the carrier density of excited carriers, respectively~\cite{LK1980}; $\alpha$ is the linewidth enhancement factor; $G_N$ represents the gain; $T_P$ and $T_S$ represent the lifetimes of photons and the inversion, respectively; $\kappa$ represents the feedback strength; $\tau_D$ represents the feedback delay; $\omega$ represents the optical angular frequency; $\varepsilon$ represents the gain saturation coefficient; and $J$ represents the injection current. $N_0$ is a constant that defines the relationship between the carrier density and the photon lifetime at the lasing threshold, as follows:
\begin{align}
    G_N(N_{th} - N_0) = \dfrac{1}{T_P},
\end{align}
where $N_{th}$ represents the carrier density at the lasing  threshold. The parameter values used in this study are presented in Table \ref{tab:parameters}~\cite{PhysRevE.86.066202}. In the numerical implementation, we utilized the fourth-order Runge--Kutta method. The timestep was 5 ps. In generating LFFs, $J$ must be set near the lasing threshold, which is determined by $J_{th} = N_{th}/T_S$. In the simulation, $J$ was fixed to $1.005 J_{th}$.

The simulation generated a 100-µs time series. Because the timestep was 5 ps, there were 20,000,000 time-series points in total.
 The entire time series was used to compute the Allan variance and PSD.

\begin{table}[htb]
    \centering
    \begin{tabular}{c|c} \hline
         Parameter & value \\\hline
         $G_N$ & $8.40\times 10^{-13} \mathrm{m^3s^{-1}}$ \\
         $N_0$ & $1.40 \times 10^{24} \mathrm{m^{-3}}$ \\
         $T_P$ & $1.927 \times 10^{-12} \mathrm{s}$ \\
         $T_S$ & $2.04 \times 10^{-9} \mathrm{s}$ \\
         $\alpha$ & $3.0$ \\
         $N_{th}$ & $2.018 \times 10^{24} \mathrm{m^{-3}}$ \\
         $\omega$ & $1.215 \times 10^{15} \mathrm{s^{-1}}$ \\
         $\varepsilon$ & $2.5 \times 10^{-23}\mathrm{m^3}$ \\
         $\tau_D$ & $2.0 \times 10^{-8} \mathrm{s}$ \\
         $J$ & $9.941 \times 10^{32} \mathrm{m^{-3}s^{-1}}$ \\ \hline
    \end{tabular}
    \caption{Parameter values in the Lang--Kobayashi equations used in the numerical simulation.}
    \label{tab:parameters}
\end{table}

\begin{figure}[htb]
    \centering
    \includegraphics[width=\columnwidth]{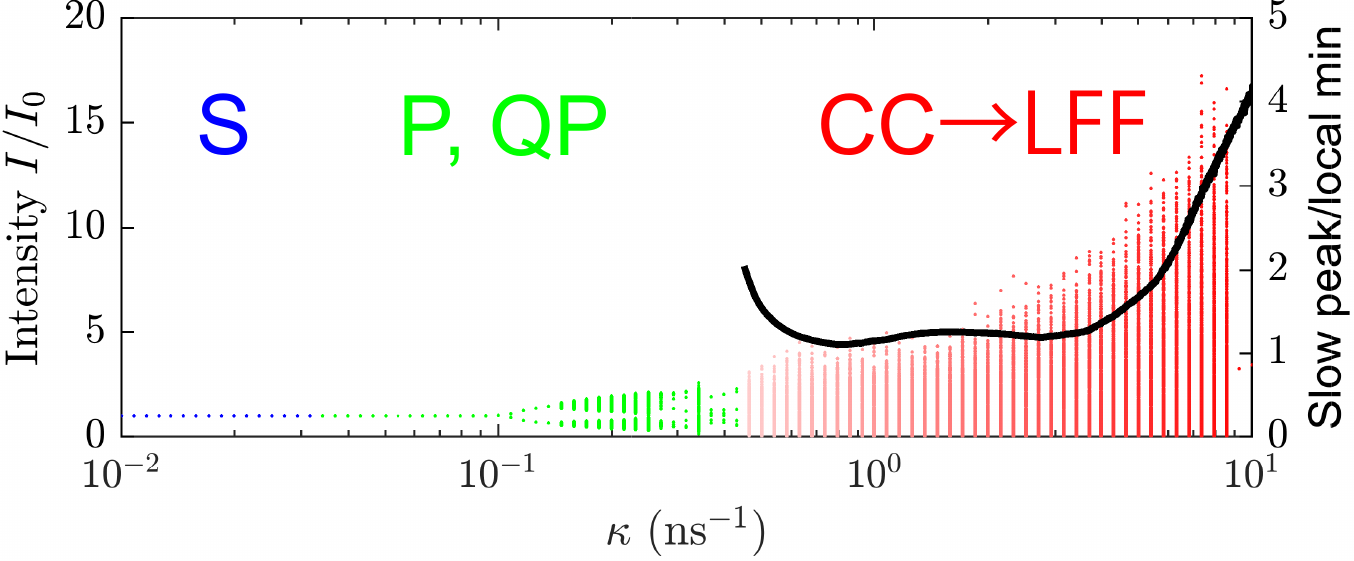}
    \caption{Bifurcation diagram of the light intensity levels $I/I_0$ with respect to the optical feedback strength indicated by the dots. As the feedback strength increases, the dynamics undergo stable emission (blue dots), periodic oscillation, quasi-periodic oscillation (green dots), CC, and LFFs (red dots). The LFFs are not clearly distinguishable from the other dynamics in the bifurcation diagram. The black curve represents a figure of merit calculated using the Allan variance. A detailed analysis is presented in Section~\ref{sec:simulation}. See also Fig.~\ref{fig:feature}.}
    \label{fig:bif}
\end{figure}

The dots in Fig.~\ref{fig:bif} show the intensity levels of the time series with respect to the feedback strength $\kappa$ in the range of $10^{-2}$ to $10$ ${\rm ns}^{-1}$. As the feedback strength increases from zero, the output of the laser undergoes constant (S), periodic (P), quasi-periodic (QP), CC, and LFF dynamics~\cite{Ohtsubo17}. However, in such a bifurcation diagram, the LFFs cannot be distinguished, because the transition to LFFs is not a bifurcation. Our objective is to systematically discriminate LFFs from other states. Meanwhile, the black line indicates the LFF decision index based on the Allan-variance analysis, which is discussed in detail in Section~\ref{sec:simulation}.

Because the transition from CC to LFF is smooth, a distinct boundary between CC and LFF cannot be immediately visualized by the indicator based on Allan variance. Therefore, the user must define a threshold appropriately depending on the target of observation. The indicator based on Allan variance varies smoothly with changes in the parameter values and can be used to classify CC and LFF once a threshold value is chosen.

\begin{figure*}[htb]
    \centering
    \includegraphics[width=1.8\columnwidth]{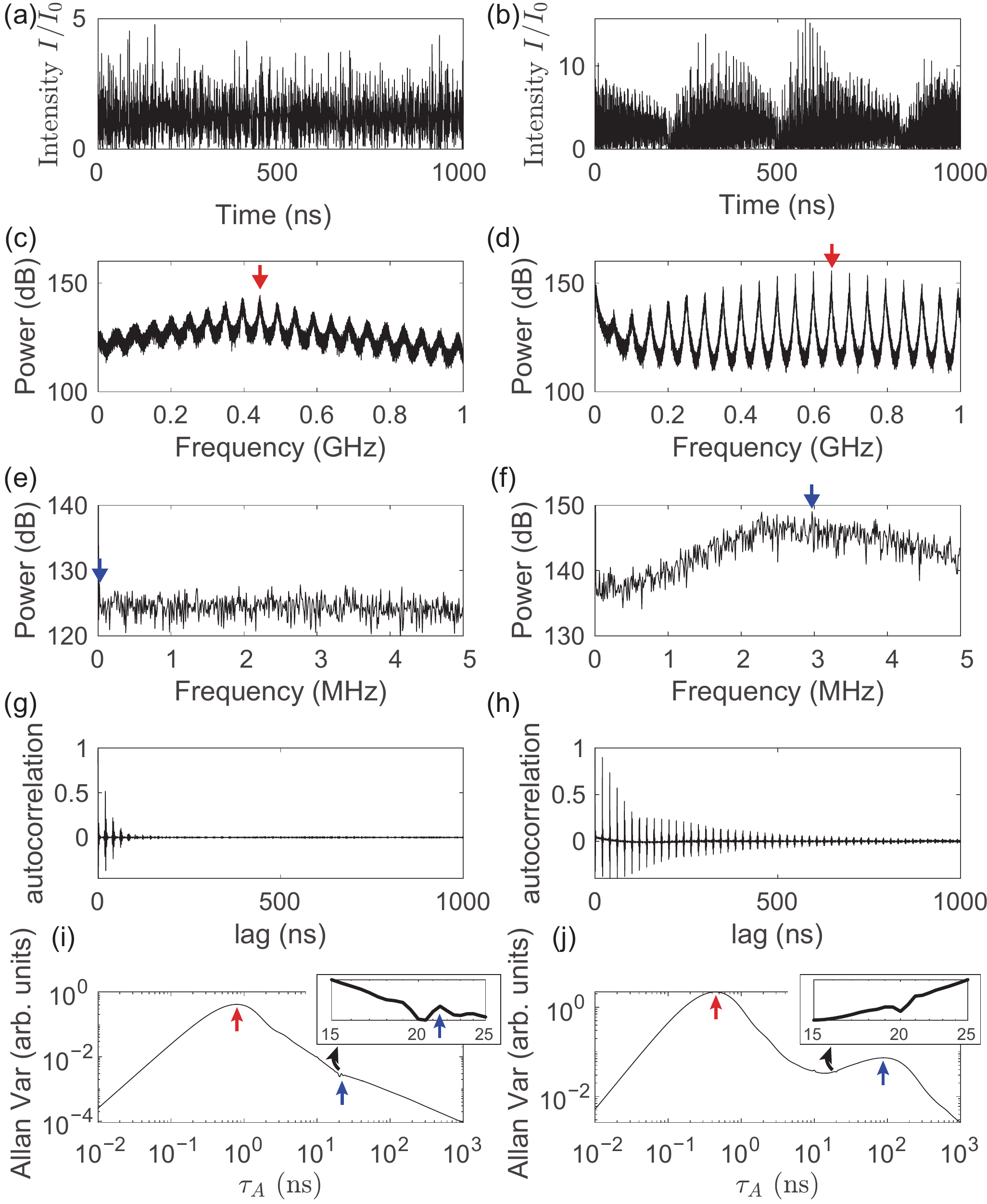}
    \caption{Comparison of CC and LFFs based on their PSDs,  autocorrelation (AC),  and Allan variance. (a, f) Time-domain snapshot, (b, g) PSD up to 1 GHz, (c, h) PSD up to 5 MHz, (d, i) AC up to 1000 ns,  and (e, j) Allan variance for time series generated with $\kappa = 1.55~{\rm ns}^{-1}$ for CC and $\kappa = 6.21~{\rm ns}^{-1}$ for LFFs. For the PSD, the red arrow indicates the global peak apart from the direct current  (DC) component, and the blue arrow indicates a local peak in the low-frequency regime apart from the DC component. The low-frequency regime is defined as 0--5 MHz. For the Allan variance, the red and blue arrows indicate the peaks in the fast and slow regimes, respectively. The slow regime is defined as $\tau_A > \tau_D$, as the timescale of dropout dynamics of LFFs are longer than $\tau_D$. Insets show magnified views of the Allan variance at approximately 20 ns, corresponding to $\tau_D$. Local minima are observed, indicating that the Allan variance can reveal $\tau_D$.}
    \label{fig:compare}
\end{figure*}

Figures~\ref{fig:compare}(a) and (f) show the time series of CC and LFFs, where the optical feedback strength $\kappa$ is 1.55 and 6.21 ${\rm ns}^{-1}$, respectively. Conventionally, the PSD has been used to characterize such time series. Figures~\ref{fig:compare}(b) and (g) present the calculated PSDs of the time series of CC and LFFs, respectively, both of which exhibit oscillatory behavior in the high-frequency regime due to the external cavity modes (ECMs)~\cite{Ohtsubo17} of the delayed feedback. 

Hereinafter, the range of 0--5 MHz in the frequency domain is referred to as the low-frequency regime, as the peak power associated with LFFs is typically observed in this regime~\cite{Uchida12}.
In LFFs, the PSD in the low-frequency regime has a high value, which stems from the sudden dropouts (Fig.~\ref{fig:compare}(h)). Comparing the magnified views of the PSD in the low-frequency regime for CC and LFFs shown in Figs.~\ref{fig:compare}(c) and (h), respectively, reveals a peak at approximately 3 MHz in the case of LFFs. However, detecting such a low-frequency peak can be technologically challenging, as mentioned in Section~\ref{sec:intro}, which is also indicated by the significant fluctuations of the PSD curves in Figs.~\ref{fig:compare}(c) and (h).

Figs.~\ref{fig:compare}(e) and (j) present the Allan variance for CC and LFFs, respectively. We observe a peak of the Allan variance at approximately $\tau_A =$ 0.5 ns in both cases, as indicated by the red arrows. For LFFs (Fig.~\ref{fig:compare}(j)), another distinct peak or local maximum is observed at approximately $\tau_A =$ 100 ns, as indicated by a blue arrow, which we refer to as the slow peak. In the following, the slow peak of the Allan variance is defined as the maximum in the slow regime, which is referred to as the region where $\tau_A > \tau_D$ in the Allan-variance plot, as the dropout dynamics of LFFs are longer than $\tau_D$. In contrast, CC (Fig.~\ref{fig:compare}(e)) does not exhibit such an evident slow peak in the slow regime.

In fact, small local minima in the Allan-variance curve are observed in Figs.~\ref{fig:compare}(e) and (j) at $\tau_A =$ 20 ns, corresponding to the round-trip time $\tau_D$ of the optical delay. These small changes are shown in the insets of Figs.~\ref{fig:compare}(e) and (j). This is one of the remarkable capabilities of the Allan variance: it can detect the signature corresponding to the round-trip delay.

A peak in the Allan variance indicates that the time series is highly variable on that timescale. The fast peaks indicated by the red arrows, which were observed for both CC and LFFs, correspond to the relaxation oscillations. In contrast, the slow peaks, which were observed only for the LFFs, capture the irregular dropout feature of the LFFs. The number of Allan-variance peaks was at most one in each of the relaxation oscillation and LFF regions. In experiments using a spectrum analyzer, there is always a possibility that the PSD peak will be missed, which is not the case with the Allan variance.

Another approach for analyzing temporal structures is the use of the autocorrelation function. Similar to the Allan variance, the autocorrelation can be directly computed from time-series data. Figs.~\ref{fig:compare}(d) and (i) show the autocorrelation functions of CC and LFF, respectively, where the latter exhibits a nonzero correlation in a larger time lag. Thus, the CC and LFFs can be distinguished via autocorrelation as well. 
However, the autocorrelation does not clearly capture the two types of timescales of LFF, i.e., the slow and fast dynamics. Rather, the round-trip time $\tau_D$ and its harmonics are clearly indicated by the peaks of the autocorrelation. From these considerations, the autocorrelation may be suited for applications such as identifying delay times in experiments. In contrast, the Allan variance is considered to be superior for capturing multiple timescales of systems from the time series observed.

\begin{figure}[htb]
    \centering
    \includegraphics[width=\columnwidth]{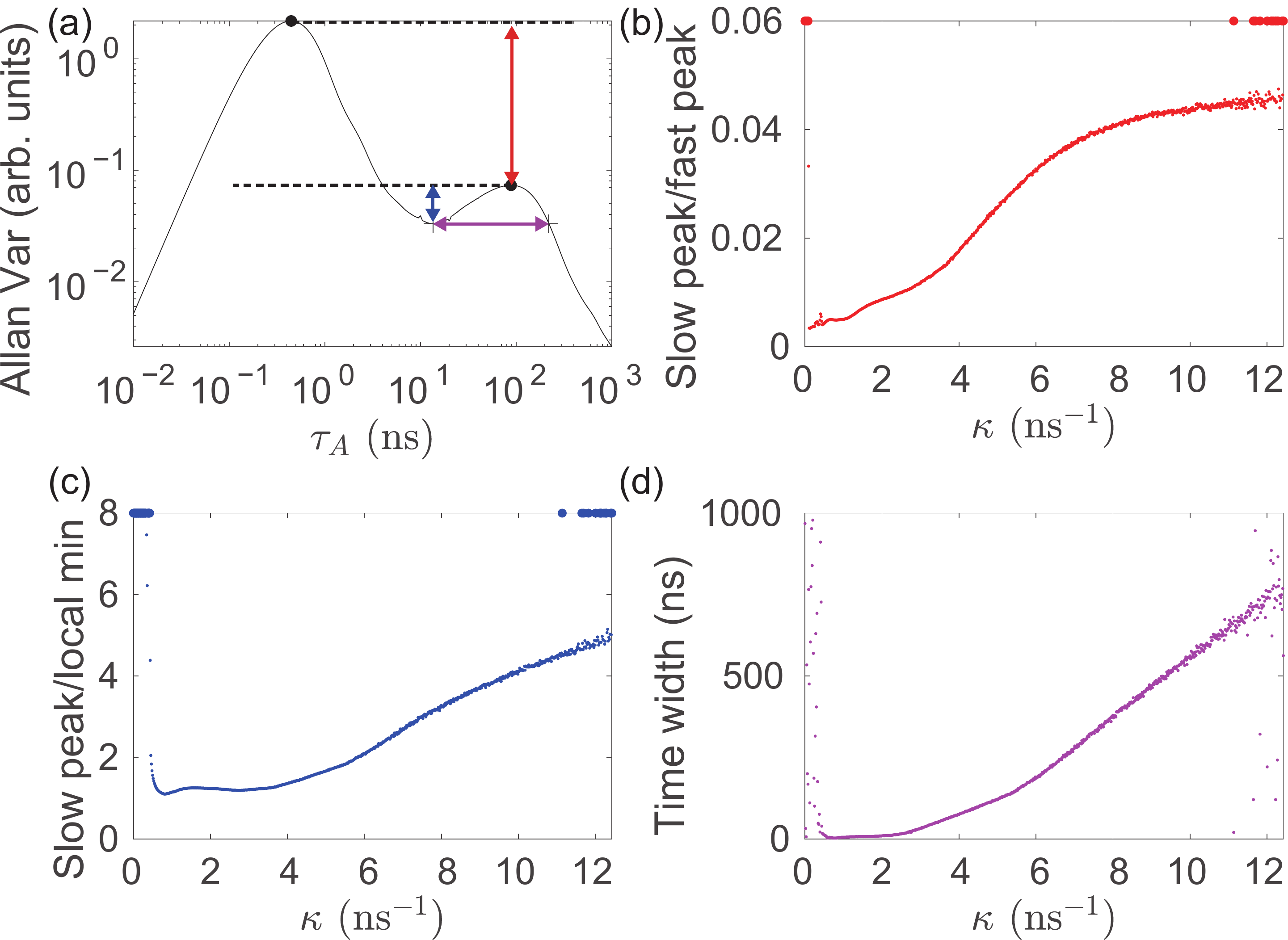}
    \caption{Characterizing fast and slow peaks in the Allan variance. (a) Representative features. The Allan-variance curve is for $\kappa = 6.21~{\rm ns}^{-1}$. (b) Ratio of the slow peak to the fast peak. (c) Ratio of the slow peak to the minimum between the two peaks. (d) Width of the slow peak. Large outliers are saturated and displayed as large dots for visibility.}
    \label{fig:feature}
\end{figure}

In Figs.~\ref{fig:compare}(e) and (j), the Allan variance exhibits smooth curves compared with the PSD. Therefore, the Allan variance provides a robust figure of merit to systematically identify LFFs according to the features associated with the peaks. Therefore, we propose the following three features for evaluating the slow peak: 
\begin{enumerate}
\item The ratio of the slow peak to the fast peak.
\item The ratio of the slow peak to the minimum between the fast and slow peaks.
\item The width of the slow peak, evaluated according to the minimum between the peaks.
\end{enumerate}
The third feature is defined as
\begin{align}
    \tau_{1} - \tau_{0} \label{eq:third}
\end{align}
where $\sigma^2_s(\tau_{0})$ represents the minimum between the fast and slow peaks, and $\tau_1$ satisfies $\sigma^2_s(\tau_{1}) = \sigma^2_s(\tau_{0})$ and $\tau_1 > \tau_0$.
These three features, i.e., figures-of-merit, were examined with respect to the feedback strength, as shown in Figs.~\ref{fig:feature}(b), (c), and~(d), respectively.

In Figs.~\ref{fig:feature}(b), (c), and~(d), several outliers (or burst-like signals) are observed with the $\kappa$ value being greater than 10 ns$^{-1}$ and less than 1 ns$^{-1}$. The underlying mechanism is explained as follows. When the feedback strength is low, periodic or quasi-periodic dynamics are observed. In such cases, the Allan variance exhibits a large number of minima, leading to a burst-like output with regard to the figure of merit defined above. (See also Fig.~\ref{fig:AV_Res}(a).)  Indeed, the Allan variance of periodic or quasi-periodic signals exhibits an oscillatory curve, which is a clear signature differentiating such periodic or quasi-periodic signals from chaotic dynamics, e.g., counting the number of minima in the Allan variance is useful for distinguishing periodic or quasi-periodic signals from chaotic ones. 

When the feedback strength is high, the outliers come from so-called periodic or stable windows, where the laser intensity oscillates with tiny amplitude near the stationary solution and sometimes even converges to the stationary solution.  Therefore, the value of the Allan variance is extremely small compared to coherence collapse and LFFs, leading to bursty behavior in the figure-of-merit through the Allan variance. Therefore, such an attribute can also be utilized to identify the underlying dynamics. 

All the indices change smoothly as the feedback intensity changes, except for the bursts, and any of them or a combination of them can be used for reliable detection of LFFs.
For example, we can define the obtained time series as exhibiting LFFs if the ratio of the slow peak to the minimum exceeds 1.5 in Fig.~\ref{fig:feature}(c).
Note that users should set the threshold considering their parameters or experimental setups.

\section{\label{sec:discussion} Discussion}
\subsection{Comparison of the PSD and Allan variance}
In this section, the PSD and Allan variance are compared with regard to the identifiability of the LFFs. In a previous study~\cite{Yamasaki:21}, a time series was defined as LFFs when the inequality
\begin{align}
    \Delta P = P_{\rm LFF}(f) - P_{\rm main}(f) \geq -15~ {\rm dB} \label{eq:LFF_PSD}
\end{align}
holds.
Here, $P_{\rm LFF}(f)$ and $P_{\rm main}(f)$ represent the spectral peak power of the low-frequency region and the global spectral peak power, respectively. 
One drawback of this definition is that estimation of the PSD is not very stable; therefore, the decision based on the inequality given by Eq.~(\ref{eq:LFF_PSD}) is unstable.

In the simulation results of Figs.~\ref{fig:compare}(b), (c), (g), and (h), the estimated PSD is not smooth with frequency changes but oscillates violently. 
This is also the case in experiments, where radiofrequency (RF) spectrum analyzers are frequently used to evaluate the PSD in this setting. 
However, the resolution of an RF analyzer is technologically limited. 
Figures~\ref{fig:compare}(b) and (g)  present several peaks associated with the ECM around the global peak. 
Depending on the resolution of the apparatus being used, the global peak may be flattened and reduced by aliasing effects. 
In particular, because the LFF is often observed when the external cavity length is relatively long, the number of ECMs increases, and the widths of adjacent peaks decrease. Therefore, this problem becomes more severe. 
Additionally, the settings of the analyzer may have to be modified for measuring the peak power in the low-frequency region, as the global peak appears at a few GHz, whereas the LFF peak usually appears at several MHz.

In contrast, the Allan variance is far more stable and makes it easier to identify LFFs because the averaging process smooths out fluctuations, as shown in Figs.~\ref{fig:compare}(e) and (j). Indeed, the proposed method based on the Allan variance can determine the LFF in a stable manner, even under the experimental constraints.

\begin{figure}[htb]
    \centering
    \includegraphics[width=\columnwidth]{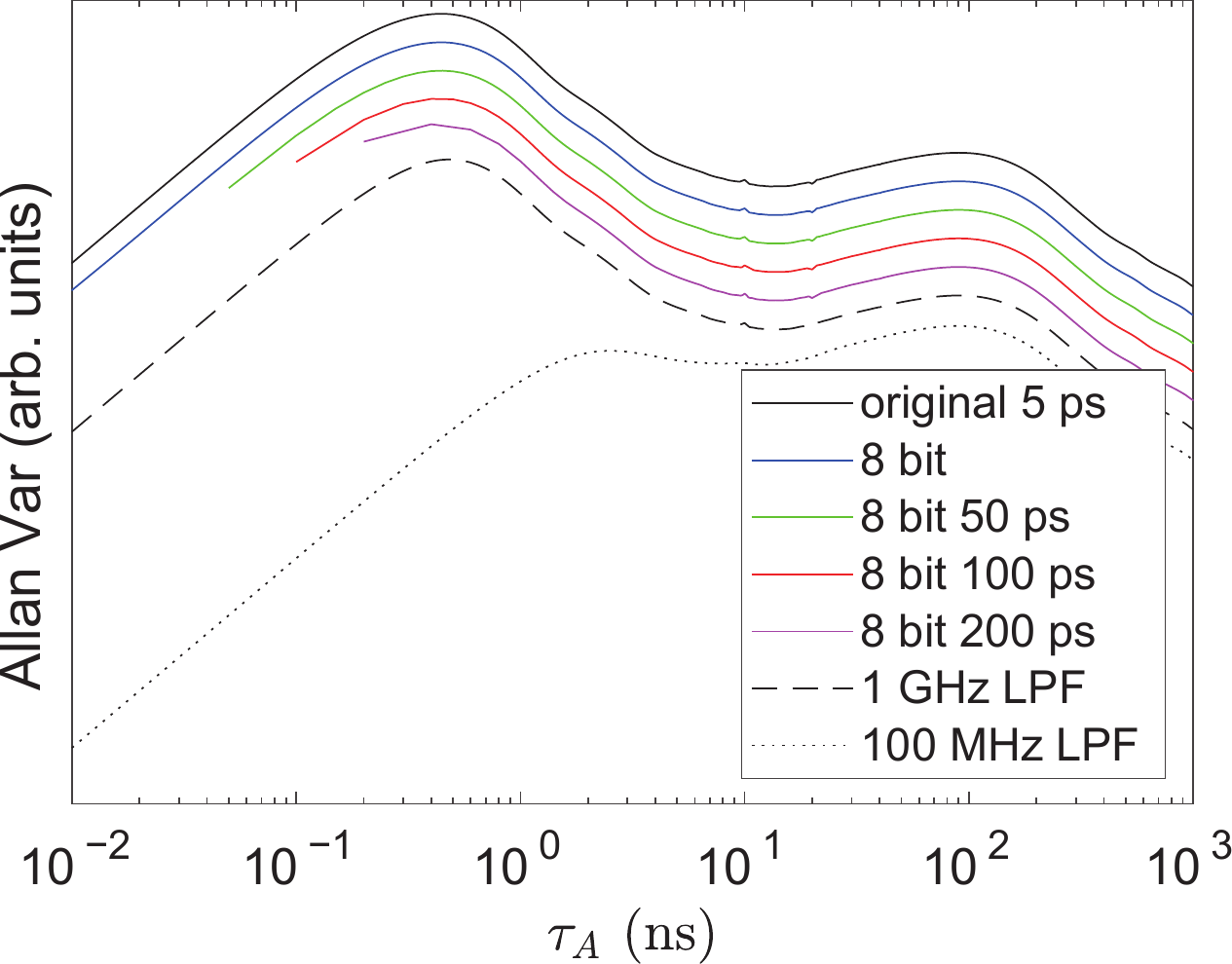}
    \caption{Allan variance of the LFF time series considering experimental limitations. The black curve is the Allan variance from the original simulation. The curve is identical to that of Fig.~\ref{fig:compare}(h). Colored curves indicate the Allan variance for 8-bit quantized and downsampled LFF time series. Dashed and dotted curves indicate the Allan variance for the low-pass-filtered LFF time series. Offsets are introduced to each curve except for the original line (black solid curve) for visibility.}
    \label{fig:quantiz}
\end{figure}

Fig.~\ref{fig:quantiz} presents the Allan variance of the LFF, taking into account the technological limitation of typical oscilloscopes with regard to data acquisition. First, the time resolution of the time series was 5 ps in the original analysis (shown in Fig.~\ref{fig:compare}(j)). 
We considered time series that are downsampled to 50, 100, and 200 ps. Second, the resolution of the signal was limited in the experimental apparatus. 
Here, we consider 8-bit quantization, i.e., 256 signal levels between the maximum and minimum values of the time series, which applies to our former experimental observation of LFF~\cite{Yamasaki:21}. 
We also assume a sufficiently long time series to prevent the total number of points from decreasing because of downsampling. After downsampling, the total number of points in the time series was 20,000,000.

The blue, green, red, and magenta curves in Fig.~\ref{fig:quantiz} show the Allan variance with an 8-bit resolution when the sampling was conducted with intervals of 5, 50, 100, and 200 ps, respectively. 
The smallest evaluable $\tau_A$ is the sampling interval.
Thus, for a coarser time resolution, a smaller $\tau_A$ become inaccessible.
In general, we see from Fig.~\ref{fig:quantiz} that as the sampling interval increased, the fast peak was slightly overestimated, while the slow peak remained almost the same. 
The difference is attributed to the downsampling, as the quantization reduced the averaging effect for small $\tau_A$ values in the fast peak, whereas the slow peak did not suffer from such effects. 
The analysis in Fig.~\ref{fig:quantiz} considers downsampling up to 200 ps to evaluate the fast peak; it would be possible to utilize an experimental apparatus with even longer sampling intervals when the main objective is the LFF determination and the detection of the slow peak only.

In addition to the sampling interval and the resolution of the signal level, the bandwidth of the data-acquisition apparatus may be a concern. To quantitatively evaluate this effect, we evaluated the Allan variance of LFF time series after infinite impulse response low-pass filtering and examined the impacts of bandwidth limitations. 
The dashed and dotted curves in Fig.~\ref{fig:quantiz} show the Allan variance when the cutoff frequency of the low-pass filter was 1 GHz and 100 MHz, respectively. 
The slow peak of LFF was successfully captured, even after the low-pass filtering, whereas the 100-MHz low-pass filter missed the fast peak. 
The 1-GHz low-pass filter changed the shape of the Allan variance, but only in the small-$\tau_A$ regime. From these results, similar to the former sampling-interval discussions, we conclude that even a small-bandwidth apparatus can detect the slow peak and is suitable in cases where accurate recognition of the fast peak is not needed. Meanwhile, typical laboratory oscilloscopes for chaotic lasers, such as those used in~\cite{Yamasaki:21}, have sufficiently large bandwidths for fast-peak characterization. 

Regarding the length of the required time series, although it is empirical, approximately 100 points of $\bar{s}_k^{(\tau_A)}$ for the largest $\tau_A$ ($\tau_A^{max}$) should be sufficient. $\tau_A^{max}\times$ sampling frequency $\times$ 100 gives the required length of the time series. In our simulation, $\tau_A^{max}$ was $10^3$ ns, and sampling frequency was 200 GHz; thus, the length of the time series was 20,000,000. However, as shown in Fig.~\ref{fig:quantiz}, a lower sampling frequency was sufficient to capture the characteristics of the slow peak, and because the peak was approximately 100 ns, a smaller $\tau_A^{max}$ could be used. Therefore, a shorter time series should be sufficient for LFF detection. The number of points can be further reduced using the refined composite method~\cite{WU20141369}, which has been proposed in the field of nonlinear time-series analysis in recent years.

\begin{figure}[htb]
    \centering
    \includegraphics[width = \columnwidth]{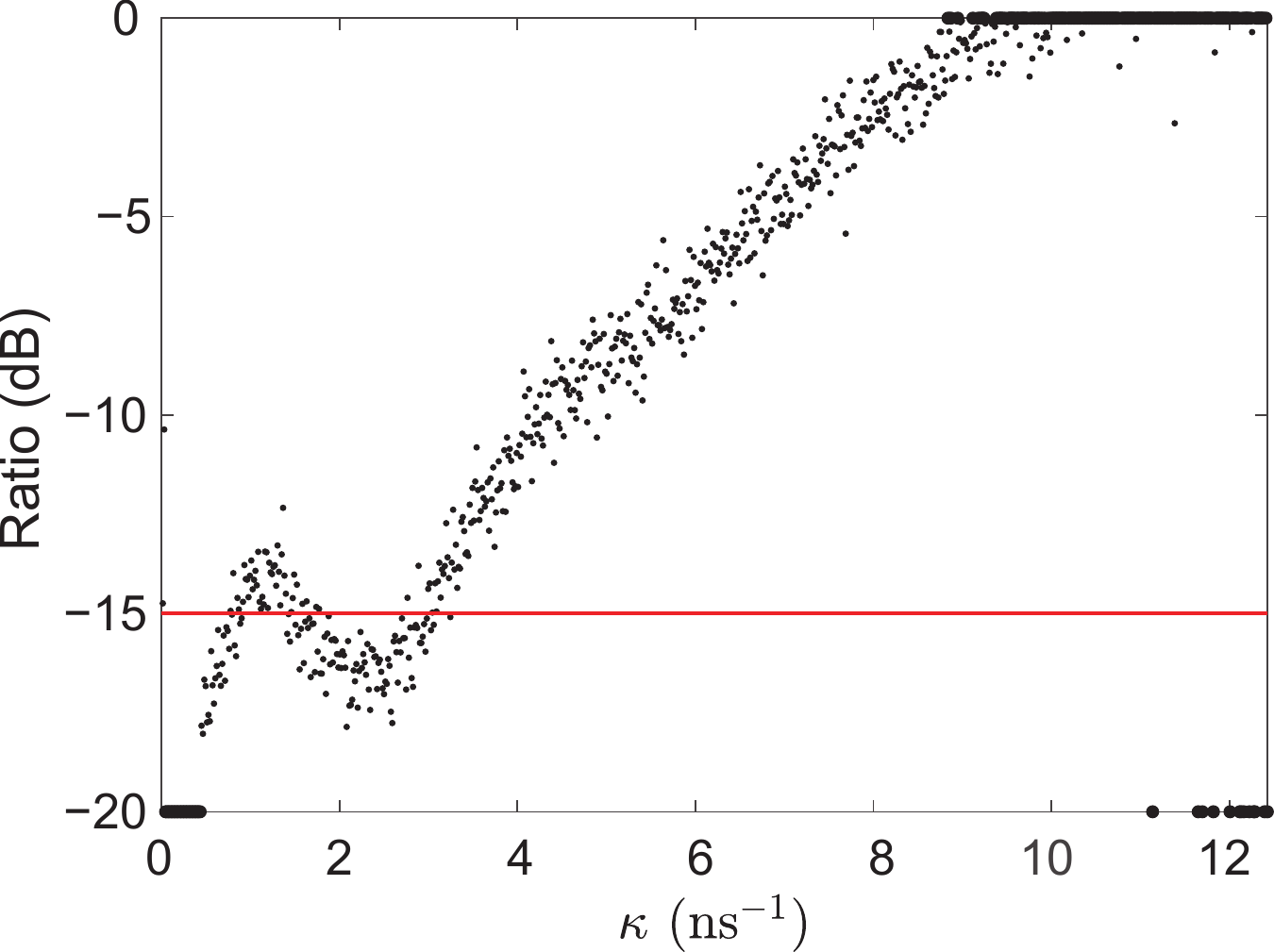}
    \caption{PSD-based characterization. The ratio of the LFF peak power to the overall peak power was evaluated as a function of feedback strength. Large outliers are saturated and displayed as large dots for visibility. In a previous study~\cite{Yamasaki:21}, when the ratio exceeded $-15$ dB, as indicated by the red line, the time-series data were classified as LFFs. The ratio was less stable and fluctuates more significantly compared to the Allan variance-based measures (see Fig.~\ref{fig:feature}).}
    \label{fig:compare_ratio}
\end{figure}

Finally, the curve in Fig.~\ref{fig:compare_ratio} shows the ratio of the LFF peak power to the overall peak power obtained via the PSD as a function of the optical feedback strength $\kappa$. 
In a previous study~\cite{Yamasaki:21}, when the ratio exceeded $-15$ dB, as indicated by the red line, the time-series data were classified as LFFs. 
The threshold value used here ($-15$ dB) was set with consideration of the device under study, which was a quantum-dot laser; hence, this threshold value may not be appropriate for other situations. 
However, regardless of the threshold level, the ratio crossed the threshold many times. Thus, it is difficult to define a specific value of $\kappa$ for discriminating different dynamics. The ratio obtained via PSD does not allow stable identification.

\subsection{Physical insights into the Allan variance peaks \label{sec:peaks}}
In this section, we discuss the physical interpretation of the Allan-variance peaks. 
The Allan variance can be expressed using the PSD; as indicated by Eq.~(\ref{eq:AV_PSD}), it is the integral of the PSD filtered with a window function, which is specified by $\tau_A$. 
Therefore, Eq.~(\ref{eq:AV_PSD}) indicates that the Allan variance is closely related to the filter function, which depends on $\tau_A$. 
When the Allan variance exhibits its peak at a certain $\tau_A$, the filter function is considered to cover the PSD in the most dominant way, i.e., that $\tau_A$ represents the dominant band. 
As mentioned previously, Eq.~(\ref{eq:AV_PSD}) indicates that the Allan variance is the PSD smoothed with the window function varying according to $\tau_A$. 

Thus, the change in the Allan variance with respect to $\tau_A$ exhibits smaller oscillations than the change in PSD with respect to the frequency. This indicates that the Allan variance is a more stable indicator than the PSD, and the LFF determination method based on Allan variance is considered to be stable.

From this perspective, the meanings of the two peaks in the Allan variance of the LFFs become clear. 
The fast peak represents the dominant band on the order of GHz, and the slow peak represents the dominant band on the order of MHz. From Eq.~(\ref{eq:AV_PSD}), we can understand why the Allan variance exhibits many minima in the case of periodic dynamics, as mentioned in Section~\ref{sec:simulation}. 
Because the filter function is proportional to the fourth power of the sinusoidal function, it goes to zero periodically. 
In contrast, when the dynamics are periodic, the PSD is spiky, and when $\tau_A$ is changed, the spikes coincide with the zeros of the filter function many times, causing the Allan variance to reach a minimum each time.

\subsection{Resolution of the Allan variance}
\begin{figure}[H]
    \centering
    \includegraphics[width=\columnwidth]{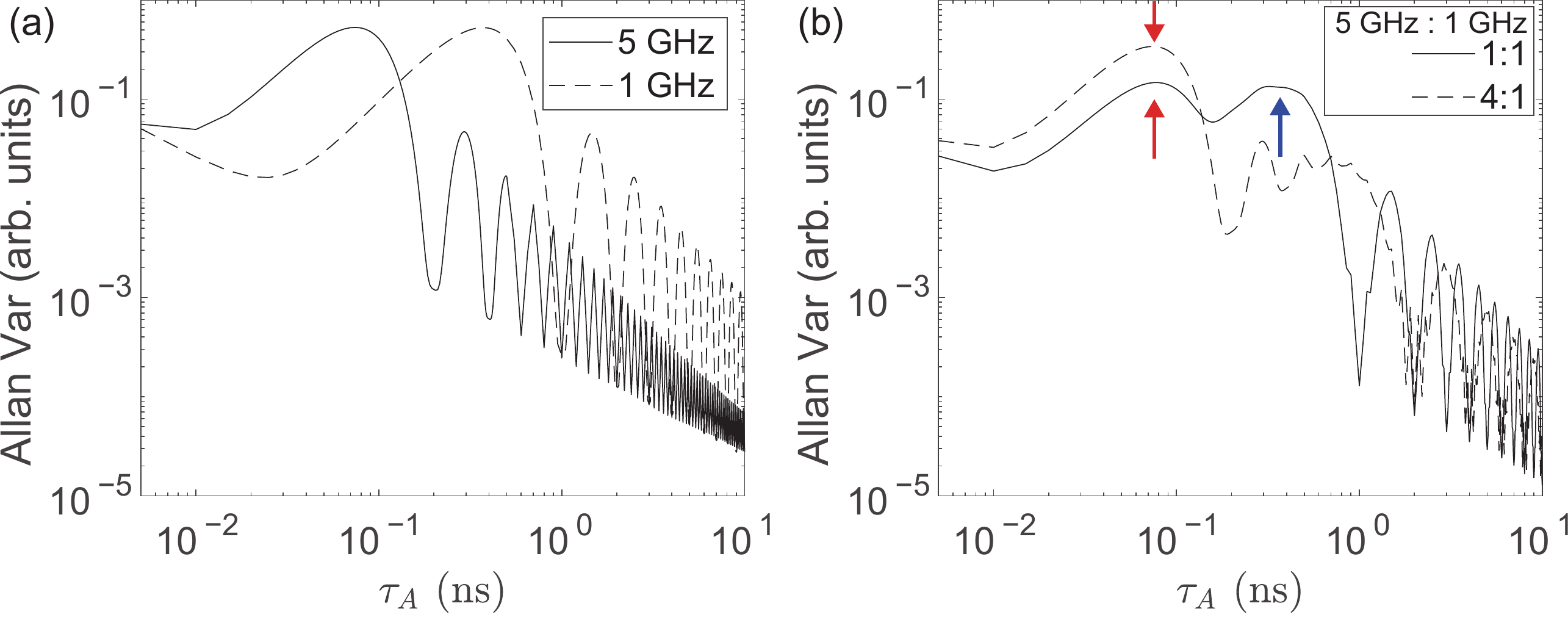}
    \caption{Resolution of the Allan variance. (a) Allan variance of the sinusoidal function with additive Gaussian noise. (b) Allan variance of sine waves of different frequencies added together at a fixed rate. The red and blue arrows indicate the peaks corresponding to the 5- and 1-GHz components, respectively.}
    \label{fig:AV_Res}
\end{figure}

The representative timescales observed for LFF, i.e., the slow and fast peaks in the Allan variance, differed by a factor of approximately 1000. In general situations, however, systems may contain numerous timescales (more than two), and these timescales may be closer. Here, we examine the resolution of the Allan variance using a simple model to clarify how closely located timescales can be distinguished. The dashed and solid curves in Fig.~\ref{fig:AV_Res}(a) indicate the Allan variance for 1- and 5-GHz sinusoids, respectively, with 10 dB of additive Gaussian noise imposed. The frequency difference between these two is significantly smaller than that in the LFF case.

The solid curve in Fig.~\ref{fig:AV_Res}(b) indicates the Allan variance for the sum of these sinusoidal signals with a ratio of 1:1. The two peaks of the Allan variance are successfully distinguished despite the small timescale difference. However, when one of the components dominates, it becomes difficult to distinguish the two types of inherent timescales according to their peaks. The dashed curve in Fig.~\ref{fig:AV_Res}(b) indicates the Allan variance when the ratio of the 5-GHz amplitude to the 1-GHz amplitude is 4:1, where the two original components cannot be distinguished. In such cases, the timescales need to be farther apart.

\section{\label{sec:conclusion} Conclusion}
We performed Allan-variance analysis of laser chaos-based complex-intensity time series for the first time. 
The results indicated that the Allan variance can be used to reliably evaluate the variability characteristics of time series over a wide range of timescales, which are manifested by a two-peak structure in the case of LFFs. 
Low-frequency components can be characterized in a stable manner compared with conventional PSD approaches. 

Finally, we discuss future prospects. In the present study, we employed a numerical approach based on the Lang--Kobayashi equations. 
Experimental approaches may enhance the practical advantages of the Allan variance over pure PSD---particularly for evaluating slow dynamics in the region from 1 to 10 Hz and below, where RF spectrum analyzers have difficulties with accurate measurements. 
Meanwhile, a deeper understanding of the multi-scale dynamics in lasers and laser networks, which may interact with external stimuli, will be important for future functional photonic devices and systems. 
Furthermore, an information theory-based understanding of multi-scale chaotic dynamics is an interesting future research direction inspired by the Allan-variance analysis of the present study. 
This study paves the way for the understanding and utilization of fast and slow dynamics and reveals how the analysis can benefit from the ideas behind the Allan variance.

\section*{Acknowledgments }
This work was supported in part by the CREST Project (JPMJCR17N2) funded by the Japan Science and Technology Agency and Grants-in-Aid for Scientific Research (JP20H00233, JP19H00868, and JP20K15185) and Transformative Research Areas (A) (22H05197) funded by the Japan Society for the Promotion of Science (JSPS). AR is supported by JSPS as an International Research Fellow.

\bibliographystyle{aip}
\bibliography{references}
\end{document}